# Emergence of a spin Hall topological Hall effect in the non-collinear phase of the ferrimagnetic insulator terbium-iron garnet

M. Loyal,[1] A. Akashdeep,[1] E. Mangini,[1] E. Galíndez-Ruales,[1] M. Eich,[1] N. Wang,[2] Q. Lan,[2] L. Jin,[2] R. Dunin-Borkowski,[2] T. Kuschel,[1] M. Kläui,[1,3] and G. Jakob[1]

[1]Institute of Physics, Johannes Gutenberg University Mainz, Staudingerweg 7, 55128 Mainz, Germany

[2]Ernst Ruska-Centre for Microscopy and Spectroscopy with Electrons (ER-C-1), Forschungszentrum Jülich GmbH, 52425 Jülich, Germany

[3]Center for Quantum Spintronics, Department of Physics, Norwegian University of Science and Technology, 7491 Trondheim, Norway

**ABSTRACT**

Magnetic compensation in rare-earth iron garnets (REIGs) offers a unique setting for which competing sublattice moments can give rise to non-collinear (canted) magnetic configurations, in which the sublattice magnetizations are not aligned with each other or with the external magnetic field. We show that this compensation regime can also host non-trivial magnetic textures. To explore this behavior, we investigated (111)-oriented epitaxial $Tb_3Fe_5O_{12}$/Pt heterostructures across the compensation temperature region using combined transverse magneto-transport and polar Kerr microscopy. Notably, we observe a topological Hall–like signal in the vicinity of the compensation temperature, a feature often interpreted as evidence for skyrmions in the absence of direct imaging. Here, in contrast, complementary Kerr microscopy reveals instead a non-collinear multidomain state which collapses outside the compensation regime, correlating directly with the appearance and disappearance of the spin Hall topological Hall effect (SH-THE) signal. These observations cannot be accounted for by a simple multi-anomalous-Hall-effect model, ruling out common artifacts as the origin, but indicate the presence of a topologically non-trivial contribution to the Hall response. These results establish strained REIG films as a tunable platform for exploring topological responses arising from compensation-driven non-collinear ferrimagnetic phases.

## I. INTRODUCTION

The topological Hall effect (THE) is widely regarded as a transport hallmark of topologically non-trivial spin textures that generate an emergent real-space magnetic field [1–5]. Consequently, hump-like anomalies in the Hall response are frequently interpreted as evidence for skyrmion formation [6–8]. Such signatures have been reported across metallic and insulating heterostructures, including rare-earth iron garnets (REIGs) coupled to heavy metals [9–12]. In the case of insulating materials, the THE has been studied with an additional heavy metal layer (e.g. Pt) that provides the ability to utilize the spin Hall magnetoresistance (SMR) [13-16] to investigate the magnetic properties of the insulator via the resistivity change in the heavy metal. For transverse SMR configuration and out-of-plane (OOP) magnetic field orientation, the spin Hall anomalous Hall effect (SH-AHE) can be measured [17]. If the insulating material possesses topologically non-trivial spin textures, the spin Hall topological Hall effect (SH-THE) might be detectable.

In numerous cases, the identification of a topological texture relies entirely on transport, with skyrmion densities inferred from the Hall signal, often without any imaging or independent proof of the underlying spin texture [6]. In systems for which imaging has been performed, the experimentally observed spin structure size frequently contradicts the skyrmion densities required to reproduce the measured Hall signal magnitude, underscoring a mismatch between transport-derived estimates and real-space magnetic configurations [7]. These inconsistencies have led to intense debate over whether the observed anomalies truly reflect real-space topology or arise from more conventional magnetic phenomena. One established alternative is the two-anomalous-Hall-effect (2-AHE) model, in which the Hall response is a superposition of distinct AHE contributions from different magnetic sublattices, interfacial regions, or domain configurations [18–20]. This mechanism can indeed explain several reports of apparent THE-like features and provides a physically meaningful description of systems with multiple magnetic components. As a result, several recent works have emphasized the need for direct complementary magnetic imaging to accompany Hall-based claims of non-trivial spin textures like skyrmions. Without such evidence, attributing Hall anomalies to topological spin textures may be misleading [21].

REIGs constitute a fascinating class of ferrimagnetic insulators whose magnetic properties arise from a three-sublattice structure: two antiferromagnetically coupled iron sublattices on the tetrahedral (*d*) and octahedral (*a*) sites, and a rare-earth sublattice on the dodecahedral (*c*) sites that is antiparallel to the net iron moment. This architecture gives rise to the magnetic compensation temperature $T_{comp}$, at which the rare-earth and iron sublattice moments cancel exactly, producing zero net magnetization. Below $T_{comp}$, the rare-earth moments dominate the ferrimagnetic order, whereas above it the iron sublattices prevail; in both regimes the system remains largely collinear [22]. Near magnetic compensation, in a non-collinear regime, the individual sublattice moments of the garnet are no longer aligned with the net magnetization. The iron moments on the octahedral and tetrahedral sites remain strongly antiferromagnetically coupled and therefore point in opposite directions, while the rare-earth moment has its own orientation set by the competition between weaker RE–Fe exchange fields and crystal-field anisotropy. The net magnetization along the external magnetic field direction is simply the vector sum of these three contributions, and this sum becomes very small near compensation. As a result, even small angular changes of the rare-earth or iron sublattices can produce large relative changes in the net moment, making the system highly sensitive to canting and the formation of non-collinear configurations [23]. Deviations from perfect collinearity have indeed been observed in multiple REIGs, including $(BiYLu)_3(FeGa)_5O_{12}$ [24], and magneto-transport measurements in compensated garnets such as InYGdIG/Pt have directly shown that transport is sensitive to these non-collinear configurations, enabling electrical detection of canted phases [23].

Adding to this complexity is the broader magnetic phase richness found in REIGs, particularly those containing rare-earth ions with strong single-ion anisotropy [25, 26]. Terbium iron garnet ($Tb_3Fe_5O_{12}$, TbIG) is a notable example: at low temperatures (below 160 K), it hosts a "double umbrella" configuration, in which the $Tb^{3+}$ magnetic moments form two distinct conical structures with different canting angles [27]. This phase emerges from the competition between the strong uniaxial crystal-field anisotropy of $Tb^{3+}$ and its exchange coupling to the Fe sublattices. While the double-umbrella phase occurs far below $T_{comp}$, its existence highlights a key point: TbIG is inherently predisposed to non-collinearity, making the high-temperature compensation regime, for which the net magnetization vanishes, a natural setting for complex multidomain and canted textures [27–30].

Here we have investigated TbIG/Pt heterostructures across the magnetic compensation region using Hall geometry magneto-transport measurements complemented by polar Kerr microscopy. We identify a distinct THE-like contribution that emerges as SH-THE signature only within a compensation-driven non-collinear multidomain regime. Interestingly, we do not observe skyrmion or bubble-type domains commonly linked to the THE; instead, the domain structure suggests the emergence of a non-collinear magnetic phase, and the magnetic field dependence of the Hall response is inconsistent with a simple 2-AHE. These observations demonstrate that compensation-enabled non-collinear textures in TbIG can generate a genuine emergent-field response due to finite scalar spin chirality [31-33], revealing a previously unexplored topological transport regime in ferrimagnetic garnets.

## II. METHODS

Epitaxial TbIG thin films with a thickness of ~9 nm were grown on single-crystal $Gd_3Ga_5O_{12}$ (GGG) (111) substrates by pulsed laser deposition (PLD) in an ultrahigh-vacuum chamber (base pressure $<2 \times 10^{-8}$ mbar). A KrF excimer laser ($\lambda = 248$ nm) with a pulse energy of 130 mJ at 10 Hz was used for ablation. The substrates were heated to 650 °C (ramp rate: 50 K min$^{-1}$) under 0.2 mbar of oxygen pressure during deposition. After growth, the films were cooled to room temperature at $-25$ K min$^{-1}$. Reflection high-energy electron diffraction (RHEED) was used to monitor surface crystallinity *in situ*. A Pt overlayer with a thickness of ~2 nm was deposited at room temperature by DC magnetron sputtering under an Ar pressure of 0.02 mbar for transverse magneto-transport measurements. To ensure a clean TbIG/Pt interface, the sample was transferred from the PLD chamber to the sputtering system in a vacuum suitcase. Crystallinity and overall growth quality were evaluated by high-resolution X-ray diffraction (HRXRD) performed on a Bruker D8 system. Atomic-scale structural analysis was conducted using an TFS Titan Spectra microscope operated at 300 kV and equipped with a high-brightness field-emission gun, probe Cs corrector, and Super-X EDX detectors [34]. High-angle annular dark-field scanning transmission electron microscopy (HAADF-STEM) images were acquired along the $[11\bar{2}]$ zone axis to characterize the film–substrate interface quality and cation ordering. Fast Fourier transforms (FFTs) of selected regions were used to confirm epitaxial alignment. Structural overlays generated in VESTA aided the identification of cation positions and the comparison with the expected garnet structure [35].

The electrical transport measurements were conducted in the GGG/TbIG/Pt heterostructure using the van der Pauw geometry, employing an RTM2 tensormeter. While we measure the transverse resistivity $\rho_{xy}$, which contains contributions from multiple mechanisms, we refer to it in the following as the Hall response. Magnetic imaging was carried out using a commercially available Evico Magnetics Kerr microscope. The system uses a blue LED illumination source and a charge-coupled-device camera, yielding a field of view of 600×400 µm². A perpendicular magnetic field was applied using the Evico OOP electromagnet capable of generating magnetic field strengths of up to ~900 mT. The sample was mounted directly on a Peltier element for temperature control, calibrated using a Pt100 resistance temperature sensor.

## III. RESULTS AND DISCUSSION

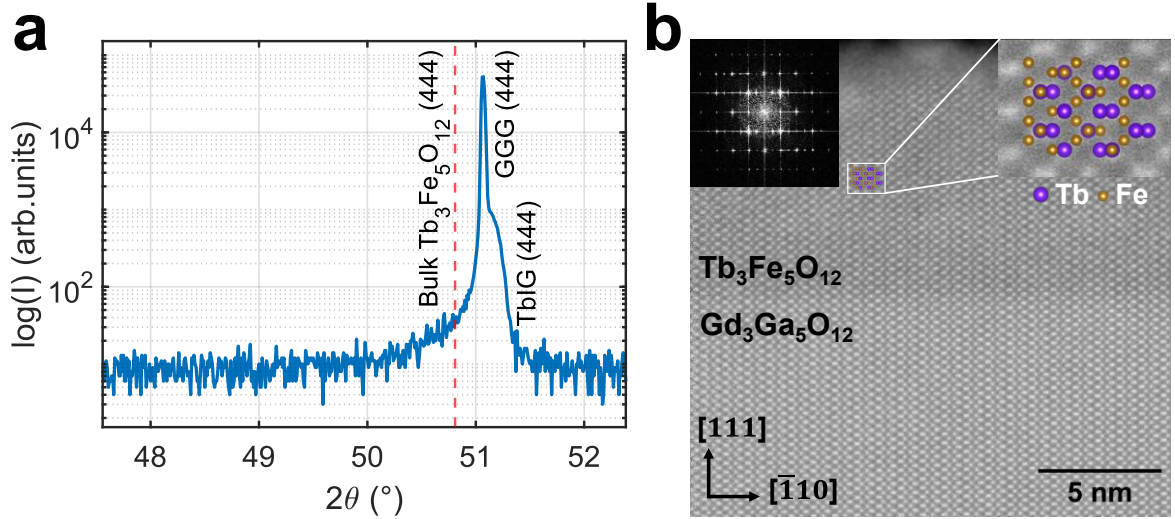

FIG. 1. Structural characterization. (a) θ/2θ HRXRD pattern for epitaxial TbIG deposited on GGG(111) substrates. Inset in the figure shows the RHEED pattern after TbIG deposition, highlighting its excellent crystalline properties. (b) HAADF-STEM image across the interface of GGG and TbIG, revealing their atomic-scale structural relationship. Top left inset shows the corresponding FFT of TbIG layer and the top right inset shows a magnified TEM region with an overlaid VESTA-generated atomic site map. The purple (Tb) and gold (Fe) markers indicate the projected crystallographic positions of the cations in the TbIG garnet structure, providing a structural reference for the atomic contrast observed in the HAADF-STEM image.

The epitaxial quality of the TbIG films is confirmed by the combined structural probes summarized in Fig. 1. The θ/2θ scan (Fig. 1(a)) exhibits a sharp GGG(444) substrate peak together with a TbIG(444) film peak that nearly coincides with it. Although bulk TbIG has a larger lattice parameter (12.44 Å) than GGG (12.36 Å), the near overlap of the peaks indicates that the TbIG thin film deviates from the bulk lattice parameter and is strongly influenced by epitaxial strain. The inset RHEED pattern further confirms the excellent epitaxial quality, revealing streaky features characteristic of a smooth, two-dimensional surface. The HAADF-STEM image acquired from a cross-sectional sample in Fig. 1(b) shows an atomically sharp interface and a well-defined garnet lattice, as highlighted by the corresponding FFT and the VESTA-generated structural overlay. Such well-defined structural quality is essential for ferrimagnetic garnets, whose magnetic compensation temperature $T_{comp}$, the point at which the net magnetization vanishes, can be strongly shifted and sensitively tuned by stoichiometry, strain, and film thickness [36−38]. Although bulk REIGs typically exhibit compensation temperatures well below room temperature, recent studies have shown that $T_{comp}$ can be substantially increased in thin films due to these effects. In the present films, the combined influence of thickness-driven strain results in a compensation temperature near room temperature (≈302 K), a regime particularly favorable for exploring interfacial and ultrafast magnetic phenomena.

### A. Compensation-driven evolution of the Hall response

The influence of the compensation point on interfacial magneto-transport is captured in the Hall response, shown in Fig. 2. At temperatures well below and above compensation, for example, at 280 K and 340 K, the Hall loops (Fig. 2(a)) are dominated by the SH-AHE, reflecting the net ferrimagnetic moment of the TbIG layer. The SH-AHE sign change across $T_{comp}$ reveals that it couples to one magnetic sublattice, reversing as the dominant contribution to the total magnetic moment shifts from the rare-earth to the iron sublattice across compensation [39]. However, as the system approaches the compensation temperature (≈302 K), the loops develop a pronounced non-monotonic feature along with a sign reversal of the SH-AHE, directly mirroring the inversion of the dominant sublattice across $T_{comp}$. The raw Hall traces and their ordinary Hall effect (OHE) fits (Fig. 2(b)) enable reliable background subtraction, and the resulting OHE-corrected curves (Fig. 2(c)) clearly resolve two distinct components: a conventional SH-AHE contribution and an additional SH-THE contribution that emerges only in a narrow temperature window around compensation. The extraction procedure is standard: linear high-field fits provide the OHE term, the SH-AHE is defined from the high-field separation of the branches, and the SH-THE component is the deviation from this OHE + SH-AHE baseline.

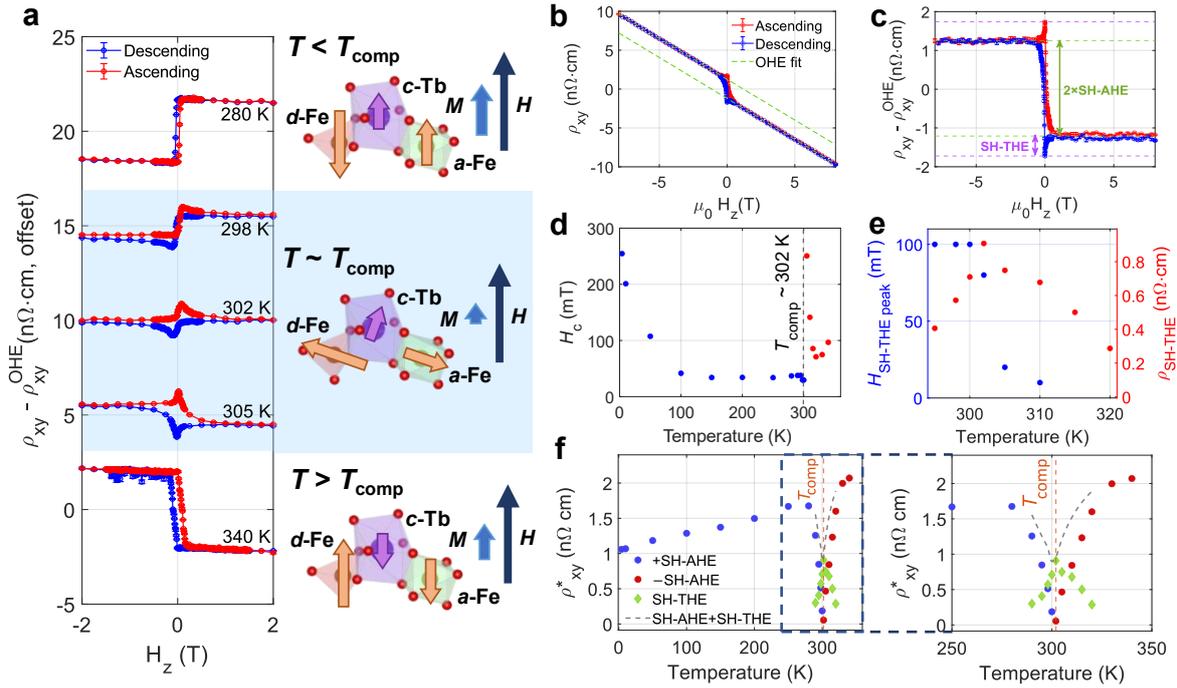

FIG.2. Temperature-dependent Hall response of GGG/TbIG/Pt heterostructure. (a) Temperature-dependent Hall resistance loops. At 280 K and 340 K, the response is dominated by a single SH-AHE contribution, whereas at temperatures around the compensation point (~302K), an additional non-monotonic feature appears together with the SH-AHE sign change across the compensation point. On the right, the corresponding configurations of the magnetic sublattices of TbIG in the non-collinear and collinear phases are shown. (b) Raw Hall data measured at 315 K with linear OHE fits used for background subtraction. (c) OHE-subtracted Hall curves revealing the SH-AHE and the additional SH-THE component at 315 K. (d) Coercivity field $H_C$ as a function of temperature for TbIG with compensation temperature ~302 K. (e) SH-THE peak amplitude ($\rho_{SH-THE}$) and corresponding magnetic field position ($H_{SH-THE\ peak}$) as a function of temperature. (f) Temperature dependence of the extracted Hall-response components $\rho^*_{xy}$ for the SH-AHE, SH-THE, and their combined signal. SH-THE is finite only near the compensation temperature, while the total SH-AHE+SH-THE curve displays an incomplete recovery, indicating the existence of a non-collinear phase.

Quantifying the temperature evolution of these components (Fig. 2(f)) shows that the coercivity diverges (Fig. 2 (d)) and SH-AHE magnitude changes sign upon crossing compensation, consistent with a reversal of the dominant sublattice. In contrast, the SH-THE contribution appearing only within a narrow temperature window (~298–320 K), peaks sharply near 302 K, and vanishes on either side of $T_{comp}$. A further indicator of intrinsic non-collinearity is the incomplete recovery of the SH-AHE component when the extracted SH-AHE and SH-THE components are recombined; the sum SH-AHE + SH-THE retains a characteristic dip. Even when the Hall loop appears saturated, the reconstructed signal does not return to the expected SH-AHE level, implying that a small degree of canting persists over an extended magnetic field range. This naturally explains why the SH-THE feature is confined to low magnetic fields while the non-collinear background survives to higher magnetic fields.

Additionally, the SH-THE peak shifts progressively toward zero magnetic field as the temperature increases above the $T_{comp}$. It reaches 0 mT between ~315 K and ~320 K before disappearing entirely (Fig. 2(e)). Any 2-AHE model, in contrast, would produce extrema at finite magnetic fields set by the sublattice switching fields. This mismatch rules out the 2-AHE explanation for this system [18]. Together, these observations indicate that ultrathin TbIG hosts a compensation-enabled realization of a magnetic state that stabilizes a non-collinear background over a broad magnetic field range, and that within this regime a distinct non-trivial spin texture emerges, giving rise to the observed SH-THE peak.

Similar Hall signal anomalies have been reported previously in TmIG [5, 6, 41], but the underlying physics there differs fundamentally from the behavior observed here. In TmIG, the SH-THE signal accompanies a transition from perpendicular to in-plane anisotropy, and its interpretation has involved a range of hypotheses, including interfacial Dzyaloshinskii-Moriya interaction (DMI)–stabilized skyrmions [6] and 2-AHE model describing from magnetic-proximity effect and spin Hall contributions [42]. Separate studies on TmIG have even reported real-space skyrmions using Kerr microscopy [9, 43], whereas in other reports only a Hall-based SH-THE signature was observed without imaging [5, 6, 41], further contributing to the diversity of interpretations. This contrast underscores the inherent ambiguity of transport-only signatures and emphasizes the need to correlate Hall measurements with direct magnetic imaging to determine the origin of SH-THE features.

### B. Real-space signatures of non-collinearity across compensation

Polar Kerr imaging provides the corresponding real-space picture, summarized in Fig. 3, and offers direct insights into the magnetic domain evolution of TbIG across the compensation region, complementing the Hall-transport signatures discussed in Fig. 2. The difference Kerr images recorded at 298 K, 303 K, and 309 K (Figs. 3(a–c)) reveal that the film cannot be fully saturated within the ±896 mT, maximum magnetic field available of the OOP coil in the Kerr-microscope. Instead, the magnetic field-driven reversal proceeds through multidomain states. The domains exhibit irregular shapes and multiple contrast levels, indicative of a non-uniform magnetic state. These complex, multidomain patterns appear precisely in the temperature window in which the SH-THE contribution is observed in transport, suggesting a strong link between the real-space magnetic configuration and the observed SH-THE in the compensation regime.

By contrast, at 337 K, well above the compensation point, the magnetic response becomes markedly less complex (Fig. 3(d)). Here, the film saturates at ~500 mT, and the reversal proceeds through conventional collinear domain switching with well-separated, uniform domains. In this regime only a single-sign SH-AHE is observed in transport, and no SH-THE features are present. The comparison across temperatures therefore establishes a direct correspondence: the emergence of multidomain state coincides with the temperature window in which the transport response becomes more complex, including a sign reversal of the SH-AHE and the appearance of an additional SH-THE contribution.

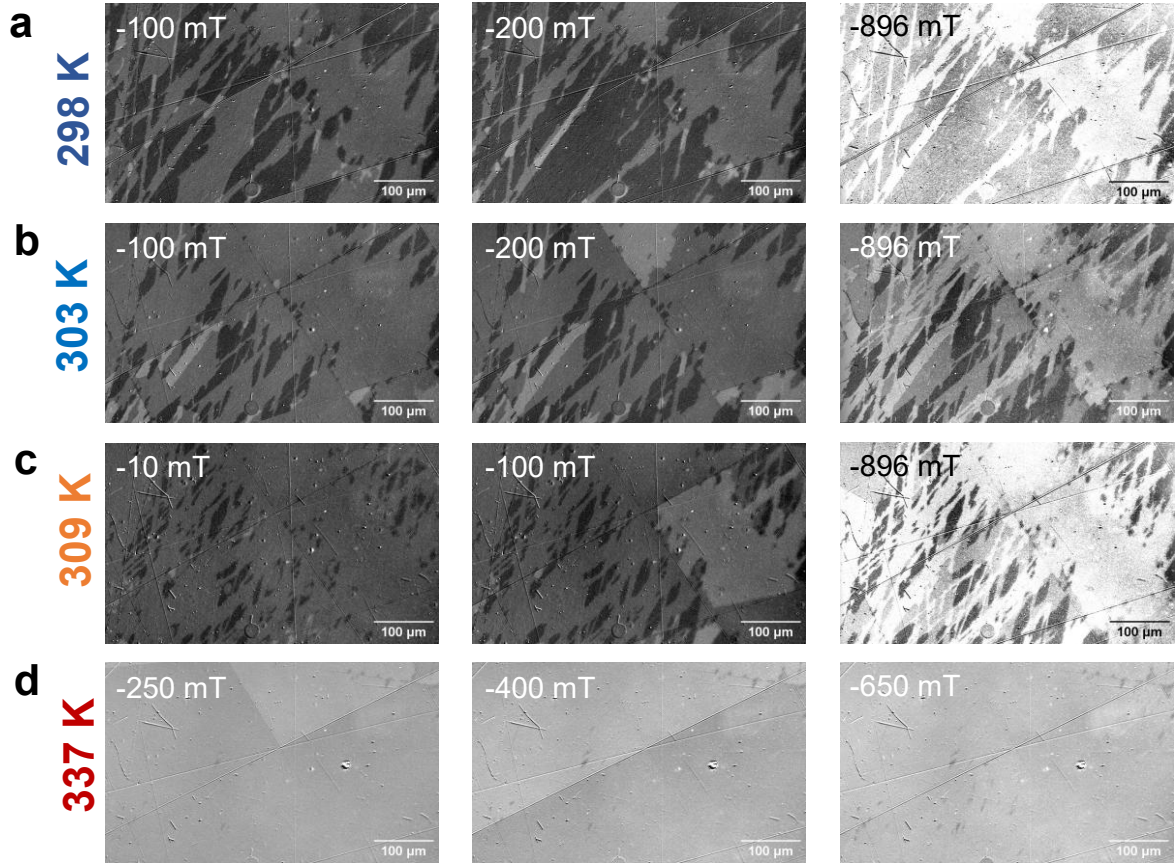

Fig. 3. Polar Kerr imaging across the compensation region. (a–d) Difference Kerr images (with respect to the +896 mT state) acquired at selected negative magnetic field strengths for four representative temperatures: 298 K (below compensation), 303 K (near compensation), 309 K (above compensation), and 337 K (far above compensation). For 298–309 K, the film cannot be fully saturated within the 896 mT OOP, and the reversal proceeds through mixed multidomain states that correspond to the SH-AHE+SH-THE features seen in transport. Around the compensation (298 K, 303 K and 309K), the domain pattern becomes irregular and multidomain, consistent with the non-collinear regime. At 337 K, for which the sample saturates at ~500 mT and only SH-AHE is observed, the reversal displays simple collinear domain switching. Magnetic field values: (a) –100, –200, –896 mT; (b) –100, –200, –896 mT; (c) –10, –100, –896 mT; (d) –250, –400, –650 mT.

### C.  Correlated imaging–transport analysis

A more detailed correlation between transport and imaging is presented in Fig. 4, which focuses on the compensation region at 315–317 K. The slight temperature offset between transport (315 K) and imaging (317 K) does not affect these conclusions; in both cases, the SH-THE feature peaks at 0 mT within the same compensation-driven non-collinear regime. The Hall signal at 315 K (Fig. 4(a)), taken on the descending branch, highlights three representative magnetic field values: (I) 0 mT: the SH-THE peak is most pronounced; (II) −60 mT: the SH-AHE switching region; and (III) −600 mT: the negative SH-AHE-saturation branch.

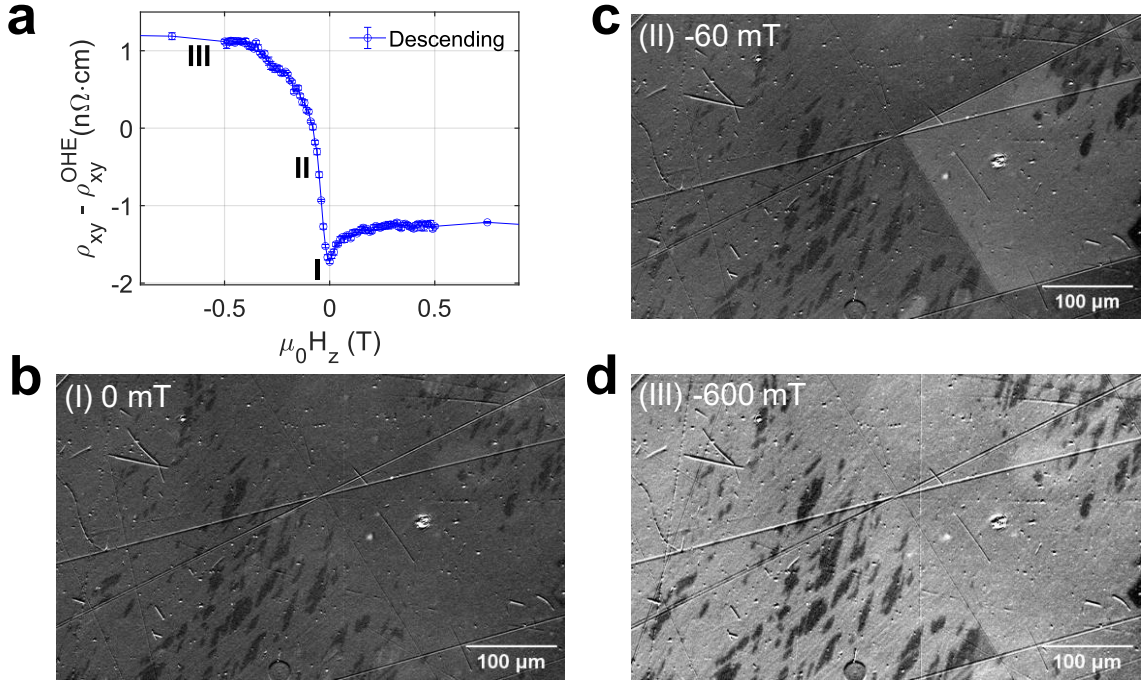

FIG. 4. Correlated Hall transport and Kerr imaging near compensation regime. (a) OHE-subtracted Hall signal measured at 315 K in the descending magnetic field branch, with three characteristic magnetic field positions marked: (I) 0 mT: maximum of the SH-THE contribution; (II) −60 mT: the SH-AHE reversal begins; and (III) −600 mT, the SH-AHE switching is completed. (b–d) Polar Kerr difference images recorded at 317 K at the magnetic field strengths indicated in panel (a). (b) At (I) 0 mT, a distinct two-level contrast with small, elongated domains appears, coinciding with the SH-THE peak. (c) At (II) −60 mT, large domains nucleate and expand, reflecting the onset of the SH-AHE-driven reversal. (d) At (III) −600 mT, the transport signal reaches the negative-saturation branch, the Kerr image still shows a multidomain state. This persistence of domains indicates that the film retains a canted, non-collinear ferrimagnetic state to magnetic field values well above the narrow window in which the SH-THE feature arises.

The Kerr images recorded at the corresponding magnetic field values (Figs. 4(b–d)) reveal how each transport feature maps onto a distinct magnetic configuration. At (I) 0 mT (Fig. 4(b)), when the SH-THE response peaks, the film exhibits a two-level, fragmented contrast composed of small, elongated domains, consistent with the non-collinear texture responsible for the SH-THE response in $\rho_{xy}$. At (II) −60 mT (Fig. 4(c)), which lies within the SH-AHE reversal window, large reversed domains begin to nucleate and expand even though the earliest switching step occurs well after the topological Hall peak vanishes. At (III) −600 mT (Fig. 4(d)), the Hall signal has reached the negative-saturation branch. Yet, the Kerr images still display residual black and white multidomain contrast at magnetic field values for which the SH-AHE signal appears saturated. While the Kerr measurement itself does not directly probe canting angles, the persistence of opposite-contrast domains at magnetic fields in which the SH-AHE signal has reached apparent saturation implies that the magnetization cannot be fully collinear. In a truly saturated (collinear) state the Kerr contrast would be uniform as in Fig. 3(d); therefore, the observed multidomain pattern is consistent with a canted magnetic configuration. A key question is whether the Hall anomaly at 315 K could arise from a trivial superposition of two anomalous Hall contributions (2-AHE), originating from two domain populations with different coercivities. In such a scenario, the additional "hump" in $\rho_{xy}$ should persist until the higher-coercivity population completes its switching. Instead, we find that the SH-THE signal collapses well before the second population reverses, which is incompatible with a simple 2-AHE superposition. The persistence of a multidomain structure well past the

SH-THE peak underscores the non-collinear nature of the TbIG layer near compensation: the non-collinear magnetic structure remains stable even under high magnetic fields, consistent with transport showing no additional switching up to 8 T (Fig. 2(b)).

Additional control measurements presented in the Supplemental Material confirm the robustness of these conclusions. In a thicker 20 nm TbIG film, the temperature interval supporting the SH-THE response collapses to a remarkably narrow ±1 K window around compensation, accompanied by a reduction in peak magnitude (see Supplemental Material, Section I). We also investigated the impact of patterning by measuring a 10 μm wide Hall bar. While the SH-THE anomaly persists, its magnetic field profile evolves from a sharp peak into a plateau-like feature. Such behavior is consistent with enhanced domain-wall pinning at device edges, which modifies the reversal dynamics without altering the fundamental compensation-driven mechanism (see Supplemental Material, Section II and III). Importantly, both the thickness-dependent and patterned-device measurements preserve the central observation: the SH-THE contribution appears exclusively within the compensation-enabled non-collinear regime and vanishes once the film re-enters a stiff collinear state. Collectively, these results and the Supplemental Material demonstrate that the emergent-field response in TbIG is robust across sample variations, while its quantitative strength and thermal bandwidth are tunable via thickness, strain, and device geometry.

## IV. CONCLUSION

In summary, we demonstrate that $Tb_3Fe_5O_{12}$ thin films host a compensation-enabled non-collinear magnetic state that can stabilize non-trivial spin textures responsible for the SH-THE signal. By combining transverse magneto-transport with magnetic field-resolved polar Kerr microscopy, we directly correlate the emergence of this transport anomaly with the appearance of a non-collinear multidomain configuration that exists only within a narrow temperature interval around magnetic compensation. Outside this regime, the domain structure reverts to conventional collinear reversal, and the Hall anomaly vanishes. The observed Hall behavior cannot be accounted for by a simple superposition of anomalous Hall components and instead points to an emergent-field Hall signal contribution driven by the compensation-enabled non-collinear, canted state exhibiting scalar spin chirality.

## ACKNOWLEDGMENTS

All authors from Mainz gratefully acknowledge funding support from the Deutsche Forschungsgemeinschaft (DFG) under the framework of the Collaborative Research Center TRR 173–268565370 Spin+X (Project B02, A01, and A12), TRR 288–422213477 Elasto-Q-Mat (Project A12). This project has also received funding from the European Research Council (ERC) under the Marie Skłodowska-Curie grant agreement No. 101119608 ('TOPOCOM') and under the European Union's Horizon 2020 research and innovation programme grant no. 856538 ('3D MAGiC'). This work contains results obtained from the experiments performed at the Ernst Ruska-Centre (ER-C) for Microscopy and Spectroscopy with electrons at the Forschungszentrum Jülich (FZJ) in Germany. The ER-C beam-time access was provided via the DFG Core Facility Project [External ER-C E-047]. M. K. acknowledges support by the Research Council of Norway through its Centers of Excellence funding scheme under project number 262633 'QuSpin'.

# Supplemental Material

# Emergence of a spin Hall topological Hall effect in the non-collinear phase of ferrimagnetic insulator terbium-iron garnet

M. Loyal,[1] A. Akashdeep,[1] E. Mangini,[1] E. Galíndez-Ruales,[1] M. Eich,[1] N. Wang,[2] Q. Lan,[2] L. Jin,[2] R. Dunin-Borkowski,[2] T. Kuschel,[1] M. Kläui,[1,3] and G. Jakob[1]

[1]Institute of Physics, Johannes Gutenberg University Mainz, Staudingerweg 7, 55128 Mainz, Germany

[2]Ernst Ruska-Centre for Microscopy and Spectroscopy with Electrons (ER-C-1), Forschungszentrum Jülich GmbH, 52425 Jülich, Germany

[3]Center for Quantum Spintronics, Department of Physics, Norwegian University of Science and Technology, 7491 Trondheim, Norway

## I. Thickness dependence

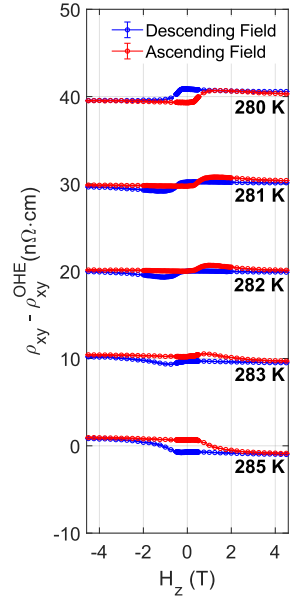

FIG. S1. **Temperature-dependent Hall response of the TbIG(20nm)/Pt heterostructure after subtraction of OHE:** At 280 K and 285 K, the response is dominated by the SH-AHE. Near the magnetic compensation temperature (~282 K), an additional non-monotonic feature associated with SH-THE appears. The SH-AHE reverses sign below and above the compensation point.

In a thicker 20 nm TbIG film, the magnetic compensation temperature shifts to a lower value of approximately 282 K, reflecting the well-known thickness dependence of magnetic compensation in REIG thin films [1]. The temperature window in which the SH-THE signal appears is sharply reduced to only about ±1 K around the compensation point (Fig. S1), and the peak magnitude is smaller than in the 9-nm film. This strong thickness dependence is consistent with the reduced influence of strain in thicker garnet layers, which stabilizes more collinear reversal and limits the formation of the non-trivial magnetic texture responsible for the SH-THE response.

## II. Geometry dependence

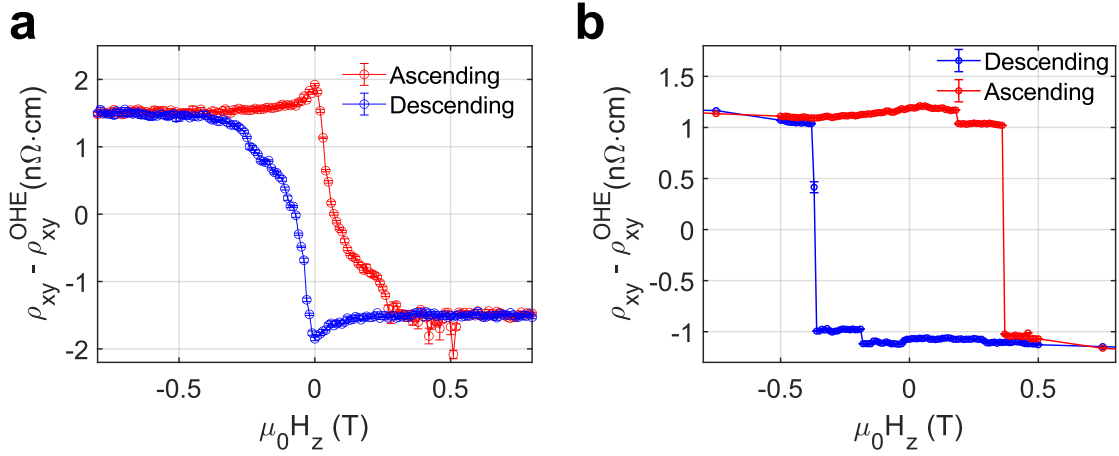

FIG. S2. Comparison of the SH-THE feature in unpatterned and patterned TbIG/Pt films. (a) OHE subtracted Hall signal from the unpatterned film showing a sharp SH-THE peak near zero field at 320 K. (b) OHE subtracted Hall signal from a lithographically defined 10 μm wide Hall bar measured at the same temperature, for which the SH-THE feature appears broadened into a plateau rather than a narrow peak. Despite the altered shape, both devices exhibit a clear non-monotonic deviation from the OHE + SH-AHE background, indicating that the emergent-field contribution persists after patterning.

Patterning the film into a micron-scale Hall bar introduces additional domain-wall pinning from lithographic edges, process-induced defects, and the confined geometry. This pinning can impede the rapid reversal of canted domains near zero field, broadening the field range over which the non-trivial spin texture persists and producing the plateau-like SH-THE feature rather than the sharp peak observed in the un-patterned film (Fig. S2). The enhanced pinning also increases the coercive field, consistent with more hindered domain-wall motion during reversal. Importantly, the raw Hall data from the patterned bar still exhibits a clear hump relative to the OHE + SH-AHE background, confirming that the underlying SH-THE response remains robust and that patterning affects only the magnetic field profile, not the presence of the underlying emergent-field-producing texture.

### III. Device fabrication and magneto-transport measurements

The as-grown samples were cleaned with acetone and isopropanol prior to patterning. The structures were patterned by photolithography employing a DMO MicroWriter ML3 and a negative photoresist, followed by $Ar^+$ ion milling for etching. Six terminal Hall bar devices of 10 μm width and 100 μm length were fabricated. For the electrical measurements, the sample was mounted with OOP magnetic field within a variable temperature insert (VTI) installed in a superconducting coil capable of generating magnetic fields up to 12 T. Measurements were conducted in a delta-mode configuration using a Keithley 6221 current source and Keithley 2182 nanovoltmeter for the Hall bar.